\documentstyle[twoside,fleqn,espcrc2,epsfig] {article}
\def\be{\begin{equation}}
\def\ee{\end{equation}}
\def\bea{\begin{eqnarray}}
\def\eea{\end{eqnarray}}

\newcommand{\AmS}{{\protect\the\textfont2
  A\kern-.1667em\lower.5ex\hbox{M}\kern-.125emS}}

\title{Atmospheric neutrino induced muons in the MACRO detector}

\author{F. Ronga (for the MACRO collaboration)
\vspace{9pt}
\\
                         INFN Laboratori Nazionali di
                         Frascati,\\
        P.O. Box 13 I-00044  Frascati Italy}%}

\begin{document}

\begin{abstract}
A measurement of the flux of neutrino-induced  muons using
the MACRO detector is presented.
Different event topologies, corresponding to different neutrino parent energies
 can be detected.
The  upward  throughgoing muon sample is the larger event sample.
For this sample, produced by neutrinos
having an average energy  around 100 GeV, the ratio of the number
of observed to expected events integrated over all zenith angles is
$0.74 \pm 0.036_{stat} \pm 0.046_{syst} \pm 0.13_{theor}$.
We have investigated whether the observed number of events and the shape
of the zenith distribution can be explained by an hypothesis of  $\nu_{\mu}
\rightarrow \nu_{\tau}$ oscillation. The best probability (17\%) is obtained
 for $\sin^2 2\theta \simeq 1.0$
and $\Delta m^2$ of a few times $10^{-3}$ eV$^2$, while the probability for
the no oscillation hypothesis is 0.1 \%.
The other samples are due to the internally produced events and to upward-going
stopping muons;
the average parent neutrino energy is  of the order of 4 GeV.
The low energy data sets  show a deficit of observed events similar to the one 
predicted by
the oscillation model with maximum mixing suggested from the upward 
throughgoing  muon sample.

\end{abstract}

\maketitle

\section{Introduction}
 The interest in precise measurements of the flux of neutrinos produced
in cosmic ray cascades in the atmosphere has been growing over the last years
due to the anomaly in the ratio of contained muon neutrino to electron neutrino
interactions.
The observations of Kamiokande, IMB and Soudan 2  are now  confirmed by
those of SuperKamiokande with larger statistics and the anomaly finds 
explanation in the
scenario of $\nu_{\mu}$ oscillations \cite{atmo}.

The effects of neutrino oscillations have to appear also in higher 
energy ranges. The flux of
muon neutrinos in the energy region from a few GeV up to a few TeV can be 
inferred from
measurements of upward throughgoing muons \cite{MACRO95}. 
As a consequence of oscillations, 
the flux of upward throughgoing muons should be affected both 
in the absolute number of events and in the
shape of the zenith angle distribution, with relatively fewer observed events 
near the vertical than
near the horizontal due to the longer path length of neutrinos from 
production to observation.

\begin{figure} [t]
\begin{center}
\epsfig{figure=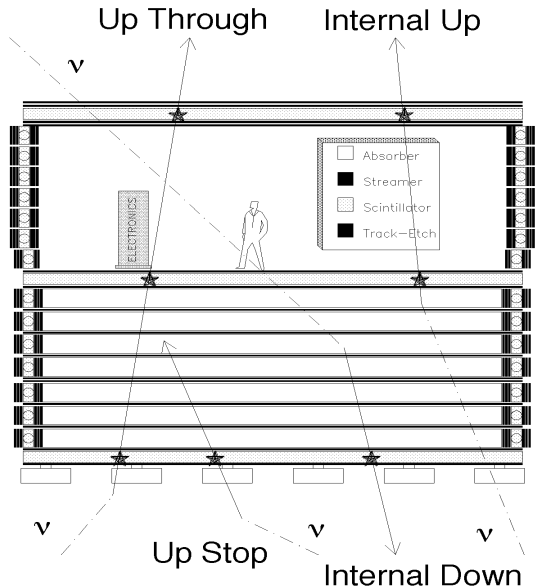,width=75mm,height=80mm} 

%\begin{figure}
%\indent{\picture 110mm by 78mm (Fig1 scaled 160)}

{\caption {\label{fig:topo}\small Sketch of different event topologies induced
by neutrino interactions in or around MACRO (see text).
In the figure, the stars
represent the scintillator hits.
The  time of flight of the particle can be measured only for the
$Internal$ $Up$ and $Up$ $Through$ events.}}
\end{center}

\end{figure}

Here the measurement about the high energy muon neutrino flux is presented, 
together with the first
results on low-energy neutrino events in MACRO.
\footnote{ The MACRO Collaboration :
M. Ambrosio,
R. Antolini,
C. Aramo,
G. Auriemma,
A. Baldini,
G. C. Barbarino,
B. C. Barish,
G. Battistoni,
R. Bellotti,
C. Bemporad
% ...
}

\section{MACRO as a neutrino detector}
The MACRO detector provides an excellent tool for the study of upgoing muons.
Its large area,
fine tracking granularity, symmetric electronics with respect to upgoing
versus downgoing muons and fully-automated analysis permit detailed studies of
the detector acceptance and possible sources of backgrounds to the
upgoing muon measurement.
In addition, the overburden of the Gran Sasso Laboratory is
significantly larger than that surrounding other
experiments (Baksan and IMB), hence
providing additional shielding against possible
sources of background induced by down-going muons.
In our first measurement of upgoing muons \cite {MACRO95},
we reported on a deficit in the total number of observed upgoing muons
with respect to the expectation and also on an
anomalous zenith angle distribution. In particular, too few muons were observed
near the nadir.
Here, we report on a much larger data set \cite{MACRO98}
which retains the same basic features as reported previously but with
larger statistics.

\begin{figure} [t]
\begin{center}
%\mbox{\hspace{-5cm}
\epsfig{file=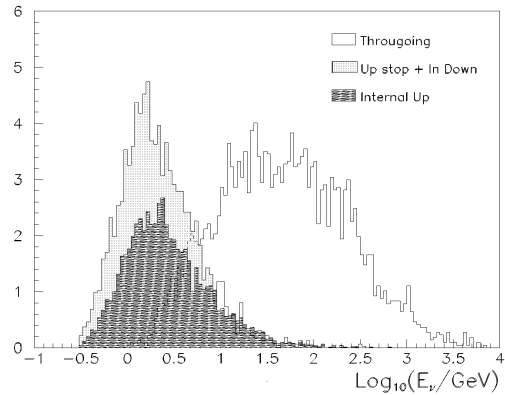,width=7.5cm,height=5.3cm}
\end{center}

%\begin{figure}
%\indent{\picture 100mm by 53mm (Fig2 scaled 160)}
{\caption{\label{fig:entopo} \small Distribution of the
parent neutrino energy giving rise to the three different
topologies of events (see Fig.~\protect\ref{fig:topo}),
computed by Monte Carlo using the
same cuts applied to the data. The distributions are normalized
to one year of data taking. The average energies of the three samples are
about 4 GeV, 4 GeV and 100 GeV, respectively.}}
\end{figure}

The MACRO detector is located in  Hall B of the Gran Sasso
Laboratory, with a minimum rock overburden of 3150 hg/cm$^2$.
It is a large rectangular box,
76.6~m~$\times$~12~m~$\times$~9.3~m, divided
longitudinally in six similar supermodules and vertically in a lower
part (4.8 m high) and an upper part (4.5 m high). The
active detection elements are planes of streamer tubes for tracking
and of liquid scintillation counters for fast timing. The lower half of
the detector is filled
with trays of crushed rock absorbers alternating with streamer tube
planes, while the upper part is open and contains the
electronics racks and work areas.  There are 10
horizontal planes in the bottom half of the detector, and 4 planes on the
top, made of wires and 27$^\circ$ stereo strip readouts.  Six vertical planes
of streamer tracking cover each side of the detector.

%The intrinsic angular resolution for muons which traverse
%the bottom of the detector is between 0.2$^\circ$
%and 1.0$^\circ$ depending on the track length. This resolution is
%smaller than the angular spread
%due to multiple scattering of muons in the rock outside of MACRO and the
%spread of angles between muons and parent neutrinos, both of which vary
%with energy.

\begin{figure}[t]
\epsfig{file=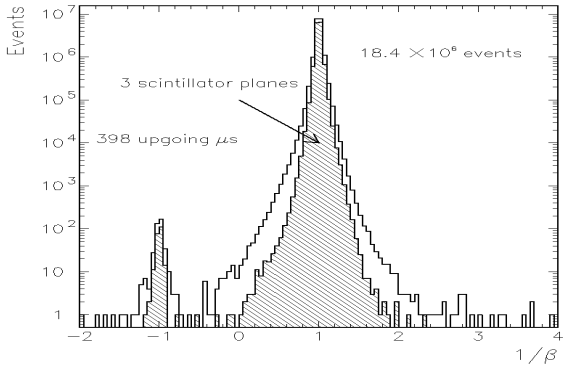,height=5.0cm,width=7.5cm}
%\end{center}
%\indent{\picture 70mm by 53mm (Fig3 scaled 265)}
\caption {\label{fig:unbeta}\small
Distribution of $1/\beta$ for the full detector data set. A clear peak of
upward muons is evident centered at $1/\beta =-1$. The widths of the
distributions for upgoing and downgoing muons are consistent. The
shaded part of the distribution is for the subset of events where
three scintillator layers were hit.}

\end{figure}

The scintillator
system consists of three layers of horizontal boxes, with vertical boxes
along the sides of the detector.  The time (position) resolution for
muons in a scintillator box in this analysis is about 500~ps ($\sim 11$~cm).

Figure~\ref{fig:topo}
shows a schematic plot of the three different topologies of neutrino 
events analyzed
up to now: $Up$ $Through$ events, $Internal$  $Up$ events and $Internal$ $Down$
together with
$Up$ $Stop$ events.  Figure~\ref{fig:entopo} shows
the parent neutrino energy distribution for the three event
topologies. The requirement of a reconstructed track selects events having 
a muon.

The  $Up$ $Through$ tracks
come from $\nu_{\mu}$ interactions in the rock below MACRO. The muon 
crosses the
whole detector ($E_{\mu}> 1$ GeV). The time information provided by
scintillator counters permits one to know the flight direction 
(time-of-flight method).
Almost 50\% of the tracks intercept 3 scintillator counters.
The average neutrino energy for this kind of events is around 100 GeV.
The data have been collected in three
periods, with different detector configurations.
In the first two periods (March 1989 - November 1991,
December 1992 - June 1993) only the lower parts of MACRO were working.
In the last  period (April 1994 - November 1997) also the upper part of MACRO 
was in acquisition.

The $Internal$ $Up$ events come from $\nu$ interactions inside the
apparatus. Since two scintillator layers are intercepted, the time-of-flight 
method is
applied to identify the upward going  events.
The average neutrino energy for this kind of events is around 4 GeV.
If the atmospheric  neutrino anomalies are the results of $\nu_{\mu}$ 
oscillations
with maximum mixing and $\Delta m^2$ between $10^{-3}$   and  $10^{-2}$ eV$^2$
it is expected a reduction
in the flux of this kind of events of about a factor of two, without any 
distortion in
the shape of the angular distribution.
Only the data collected with the full MACRO (live-time
around 3  years) have been used in this analysis.

The $Up$ $Stop$ and the  $Internal$ $Down$ events  are due to  external 
interactions with
upward-going tracks stopping in the detector ($Up$ $Stop$) and to  neutrino 
induced downgoing
tracks with vertex in lower part of MACRO ($Internal$ $Down$). These events are
identified by
means of topological criteria. The lack of time information prevents  
distinguishing the two
sub samples. The data set used for this analysis is the same used for the 
$Internal$ $Up$ search.
 An almost equal number of $Up$ $Stop$ and $Internal$ $Down$ is expected if
neutrinos do not oscillate. The average neutrino energy for this kind of events
is
around 4 GeV. In case of oscillations we expect a reduction in the flux of the
$Up$ $Stop$ events similar to the one expected for the $Internal$ $Up$ events,
while we do not expect any reduction
of the
$Internal$ $Down$ events (having path lengths of the order of 20 km).

\section{Upward throughgoing muons ($Up$ $Through$) }

The direction that muons travel through MACRO is determined by the
time-of-flight between two different layers of scintillator counters.
 The measured muon velocity is calculated with the
convention that muons going down through the detector are expected to have
1/$\beta$ near +1 while muons going up through the detector are expected
to have 1/$\beta$ near -1.

Several cuts are imposed to remove backgrounds caused by radioactivity in near
coincidence with muons and showering events which may result in bad time
reconstruction. The most important cut requires that the
position of a muon hit in each scintillator as determined from the
timing within the scintillator counter agrees
within $\pm$70 cm with the position
indicated by the streamer tube track.

When a muon hits 3 scintillator layers, there is  redundancy in the
time measurement and 1/$\beta$ is calculated from a linear fit of the times
as a function of the pathlength. Tracks with a
poor fit are rejected. Other minor cuts are applied for
 the tracks with only two layers of scintillator hit.

It has been observed that downgoing muons which pass near or through
MACRO may produce low-energy, upgoing particles.  These could appear to be
neutrino-induced
upward throughgoing muons if the down-going muon misses the detector 
\cite{BACKSCA}.
In order to reduce this background, we impose a cut
requiring that each upgoing muon must cross at least 200 g/cm$^2$ of
material in the bottom half of the detector. Finally, a large number
of nearly horizontal ($\cos \theta > -0.1$), but upgoing muons have
been observed coming from azimuth angles (in local coordinates) from
-30$^\circ$ to 120$^\circ$. This direction contains  a cliff in
the mountain where the overburden is insufficient to remove
nearly horizontal, downgoing muons which have scattered in the mountain
and appear as upgoing. We exclude this region from both our
observation and Monte-Carlo calculation of the upgoing events.

\begin{figure} [t]
\begin{center}
%\mbox{
\epsfig{file=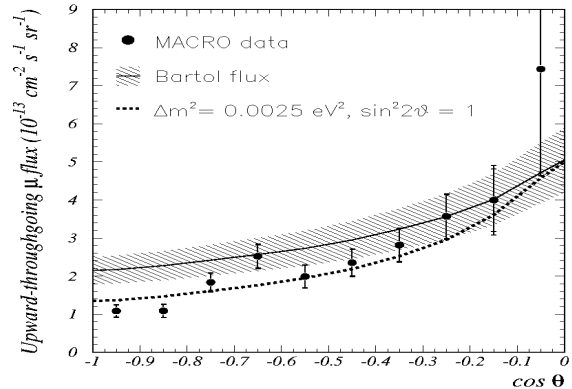,width=7.5cm,height=5.3cm}
%\indent{\picture 75mm by 53mm (Fig4 scaled 255)}
\caption {\label{angleflux}\small Zenith distribution of flux of upward 
throughgoing muons with energy
greater than 1 GeV for data and Monte Carlo for the combined MACRO
data. The solid curve shows the expectation for no oscillations and
the shaded region shows the 17\% uncertainty in the expectation. The
dashed line shows the prediction for an oscillated flux with
$\sin^2 2 \theta = 1$ and $\Delta m^2 = 0.0025$ eV$^2$.}
\end{center}
\end{figure}

Figure~\ref{fig:unbeta} shows the $1/\beta$ distribution for the
up-throughgoing data from the full detector running.
%(for the older data see the
% equivalent figure in ref. \cite {MACRO95}).
 A clear peak of upgoing
muons is evident centered on $1/\beta=-1$.
There are 398 events in the range $-1.25 < 1/\beta < -0.75$ which we
define as upgoing muons for this data set. We combine
these data with the previously published data \cite{MACRO95} (with 4 additional
events
due to an updated analysis) for a total of 479 upgoing events.
Based on events outside the upgoing muon peak, we estimate
there are $9 \pm 5$ background events in the total data set.
In addition to these events, we estimate
that there are $8 \pm 3$ events which result from upgoing
charged particles produced by
downgoing muons in the rock near MACRO.
Finally, it is estimated that $11 \pm 4$ events are the
result of interactions of neutrinos in the very bottom layer of MACRO
scintillators. Hence, removing the backgrounds, the
observed number of upgoing throughgoing muons integrated over all
zenith angles is 451.

\begin{figure} [t]
%\begin{center}
%\mbox{
\epsfig{file=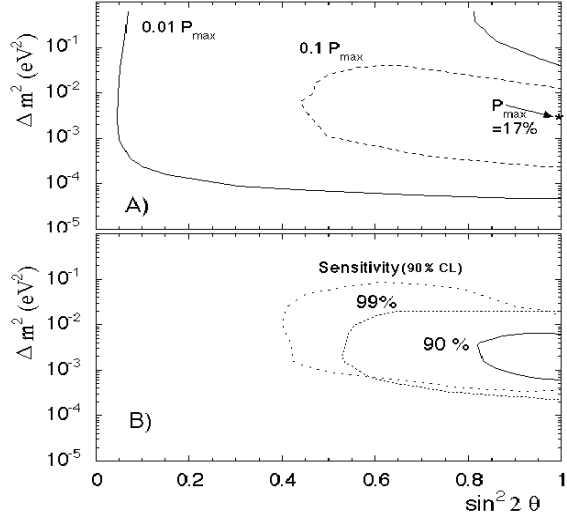,width=7.5cm,height=7.0cm}
%\indent{\picture 100mm by 73mm  (Fig5 scaled 345)}
%\end{center}
\caption
{\label{fig:exclusion}\small
A) Probability contours
  for oscillation parameters for $\nu_\mu \rightarrow \nu_\tau$
  oscillations based on the combined probabilities of zenith shape and
  number of events tests.
   The best probability in the physical regn is 17\% and iso-probability
   contours are shown for 10\% and 1\% of this value (i.e. 1.7\% and
   0.17\%).
%\\ 
B)  Confidence regions
 at the 90\% and 99\% levels calculated according to reference
   \protect\cite{Feldman-Cousins}.
   Since the best probability is outside the physical region
   the confidence intervals regions are smaller than the one expected
   from the sensitivity of the experiment.}
\end{figure}

\begin{figure}
\begin{center}
\epsfig{file=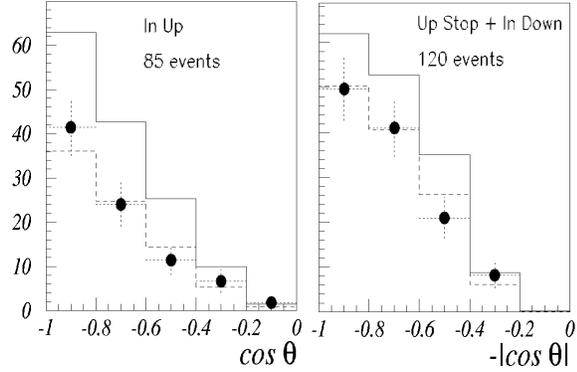,width=7.5cm,height=5.0cm}
%\mbox{
%\indent{\picture 100mm by 50mm  (Fig6 scaled 180)}
\caption{\label{fig:IUP}\small
Comparison between measured and expected number of low energy  events versus
$cos(\theta)$. The dashed line is obtained assuming neutrino oscillation with 
the parameter
suggested by the $Up$ $Through$ Sample. Note that in the second plot the flight
direction is unknow}
\end{center}
\end{figure}

In the upgoing muon simulation we have used the neutrino flux
computed by the Bartol group \cite{Agrawal96}.
 The cross-sections for the neutrino interactions have been calculated using the
 Morfin
and Tung parton distributions set S$_1$  \cite{Morfin91}. These
parton distributions were chosen based on the good agreement of the resulting
$\sigma_T$ compared to the world average at $E_\nu = 100$~GeV.
The propagation of muons to the detector has been done using
the energy loss calculation by Lohmann et al.
\cite{Lohmann85} for standard rock.
The total systematic uncertainty on the expected flux of muons
 adding the errors from neutrino flux, cross-section
and muon propagation in quadrature is $\pm17\%$.
This theoretical error in the prediction is mainly a scale error that 
doesn't change the shape of the angular
distribution.
The number of events expected integrated over all zenith angles is
612, giving a ratio of the observed number of events to the expectation
of 0.74 $\pm0.036$(stat) $\pm0.046$(systematic) $\pm0.13$(theoretical).

Figure \ref{angleflux} shows the zenith angle distribution of the measured
flux of upgoing muons with energy greater than 1 GeV for all MACRO data
compared to the Monte Carlo expectation for no oscillations
and with a 
$\nu_{\mu} \rightarrow \nu_{\tau}$ oscillated flux with 
$\sin^2 2 \theta = 1$ and
$\Delta m^2 = 0.0025$ eV$^2$ (dashed line).

%The range for the Monte Carlo expectation
%for the unoscillated flux reflects the  $\pm17\%$
%systematic uncertainty in that prediction.

The shape of the angular distribution has been tested with the hypothesis of no
oscillation
excluding the last bin near the horizontal and normalizing data and predictions.
The $\chi^2$ is $26.1$,  for 8 degrees of freedom
(probability of 0.1\% for a shape at least this different from the
expectation). 
We have considered also oscillations $\nu_{\mu} \rightarrow\nu_{\tau}$.
The best $\chi^2$ in the physical region of the oscillations parameters 
is 15.8 for $\Delta m^2$ around $0.0025eV^2$ and maximum mixing (the  best
 $\chi^2$ is outside the physical region for mixing $>1$ ).

To test oscillation hypothesis, we calculate the independent probability for
obtaining the number of events observed and the angular distribution
for various oscillation parameters. 
It is notable that the value of $\Delta m^2$
suggested from the shape of the angular distribution is similar to the value 
necessary in order
to obtain the observed reduction in the total number of events in the 
hypothesis of maximum mixing.
Figure \ref {fig:exclusion} A) shows probability contours for oscillation
parameters using the combination of probability for
the number of events and $\chi^2$ of the angular distribution.
The maximum of the probability is 17\%.
The probability for no oscillation is 0.1\%.
Figure \ref {fig:exclusion} B) shows the
confidence regions at the 90\% and 99\% confidence levels based on
application of the Monte Carlo prescription of reference \cite
{Feldman-Cousins}.  We plot also  the
sensitivity of the experiment. The sensitivity is the 90\% contour which 
would result from the preceding
prescription if the data and Monte Carlo happened to be in perfect agreement at
the
best-fit point. The allowed regions are smaller than the one you could
expect form the sensitivity of the experiment. This is because the best probabil
ity is outside the physical region.

The same procedure applied for sterile neutrino \cite{Smirnov97} gives 2\% as 
maximum probability.

\begin{table*}[hbt]
% space before first and after last column: 1.5pc
% space between columns: 3.0pc (twice the above)
\setlength{\tabcolsep}{1.5pc}
% -----------------------------------------------------
% adapted from TeX book, p. 241
\newlength{\digitwidth} \settowidth{\digitwidth}{\rm 0}
\catcode`?=\active \def?{\kern\digitwidth}
% -----------------------------------------------------
\caption{Event Summary. The predictions with oscillations are  for maximum 
mixing and $\Delta m^2=0.0025 eV^{2}$}
\label{tab:summary}
\begin{tabular*}{\textwidth}{@{}l@{\extracolsep{\fill}}rrrr}
\hline
                 & \multicolumn{1}{l}{Events detected}
                 & \multicolumn{2}{l}{Predictions (Bartol neutrino flux)} \\
\cline{3-4}
               &  & \multicolumn{1}{r}{No Oscillations}
                 & \multicolumn{1}{r}{With oscillations}         \\
\hline
$Up$ $Through$    & $ 451$  & $612\pm104_{theoret}\pm37_{syst}$ & 
$431\pm73_{theoret}\pm26_{syst}$
\\
$Internal$ $Up$   & $ 85$  & $144\pm36_{theoret}\pm14_{syst}$  & 
$83\pm21_{theoret}\pm8_{syst}$
\\
$In$ $Down$ + $Stop$ & 120 & $159\pm40_{theoret}\pm16_{syst}$ & 
$123\pm31_{theoret}\pm12_{syst}$ \\
\hline
%\multicolumn{5}{@{}p{120mm}}

\end{tabular*}
\end{table*}

\section{The Low Energy Events}
The analysis of the $Internal$ $Up$ events is  similar to the analysis of the
$Up$ $Through$. The main difference is due to the requirement that the 
interaction vertex should be inside
the apparatus. About 87\% of events are estimated to be  $\nu_{\mu}$
interactions.  The preliminary uncertainty due to the acceptance and analysis 
cuts is  10\%.
After the background subtraction (3 events) 85 events are classified as 
$Internal$ $Up$ events

The  $Internal$ $Down$ and the $Up$ $Stop$  events are identified via 
topological
constraints.  The main requirement is the presence of a reconstructed track 
crossing the
bottom scintillator layer. All the track hits must be at least 1 m  from 
the detector's edges.
The criteria used
to verify that the event vertex (or  stop point) is inside the detector 
are similar to those used for the
$Internal$ $Up$ search.  To reject ambiguous and/or
wrongly tracked events which survived automated analysis cuts, 
real and simulated events were
randomly merged and directly scanned with the MACRO Event Display. 
Three different events
subsamples are considered according to the minimum number  of streamer 
tube hits. We present
here  the sample with at least 3  streamer hits
(corresponding roughly to  100 $gr$ $cm^{-2}$). 
About 90\% of the events are estimated to be
 $\nu_{\mu}$ CC interactions. 
The main background for this search are the low energy
particles produced by donwn-going muons \cite{BACKSCA}. After  background
subtraction (5 events) 120 events are classified as $Internal$ $Down$ and  $Up$
$Stop$ events.

The Montecarlo simulation for the low energy events uses the Bartol neutrino 
flux \cite{Agrawal96}
and the neutrino low energy cross sections reported in \cite{Lipari94}.
The simulation is performed in a large volume of rock (170 kton) around the 
MACRO detector (5.3
kton).   The uncertainty on the expected muon flux is
about 25\%.
The total number of events and the angular distributions are compared with 
the predictions in
Table~\ref{tab:summary} and
in Figure~\ref{fig:IUP}. The low energy samples show an uniform deficit of 
the measured number
of events over the whole angular distribution with  respect to the predictions,
while there is a
good agreement with the predictions based on neutrino oscillations.

\section{Conclusions}
The upgoing throughgoing muon data set is in favor of 
$\nu_{\mu} \rightarrow\nu_{\tau}$
oscillation  with  parameters similar  to the those observed by Superkamiokande
with a probability of 17\% against the 0.1\% for the no
oscillation hypothesis.
However the shape of the zenith distribution  gives  a maximum
probability of only 4.6\%. 
This could be due  to a statistical fluctuation or to some hidden
physics.  We exclude effects due to the detector.

This oscillation  hypothesis is also consistent with the MACRO low energy
data. A combined statistical analysis of the three data samples will 
be performed in the future when more statistics will be available.

\end{document}